%% file: main.tex
\documentclass[
reprint,
nofootinbib,
superscriptaddress,
 amsmath,
 amssymb,
 aps,
 prapplied,
 floatfix,
 fleqn,
]{revtex4-2}
\usepackage[colorlinks=true,citecolor=blue,filecolor=blue,linkcolor=blue,urlcolor=blue]{hyperref}
\usepackage[dvipsnames]{xcolor}
\usepackage{graphicx}
\usepackage{multirow}
\usepackage{lineno}
\usepackage{tabularx}
\usepackage{hyphenat}
\usepackage{mathtools, cuted} 

\begin{document}

\title{Physics-informed machine learning approaches to reactor antineutrino detection}

\include{affiliations}
\include{authors}
\noaffiliation

\date{\today}
\begin{abstract}
\noindent 
Nuclear reactors produce a high flux of $MeV\!-$scale antineutrinos that can be observed through inverse beta-decay (IBD) interactions in particle detectors. 
Reliable detection of reactor IBD signals depends on suppression of backgrounds, both by physical shielding and vetoing and by pattern recognition and rejection in acquired data. 
A particularly challenging background to reactor antineutrino detection is from cosmogenically-induced fast neutrons, which can mimic the characteristics of an IBD signal.
In this work, we explore two methods of machine learning -- a tree-based classifier and a graph-convolutional neural network -- to improve rejection of fast neutron-induced background events in a water Cherenkov detector. 
The tree-based classifier examines classification at the reconstructed feature level, while the graphical network classifies events using only the raw signal data. 
Both methods improve the sensitivity for a background-dominant search over traditional cut-and-count methods, with the greatest improvement being from the tree-based classification method. 
These performance enhancements are relevant for reactor monitoring applications that make use of deep underground oil-based or water-based kiloton-scale detectors with multichannel, PMT-based readouts, and they are likely extensible to other similar physics analyses using this class of detector.  

\end{abstract}
\maketitle

\section{Introduction}\label{sec:introduction}

The beta-decay of fission daughter particles in operating reactors cores generates enormous numbers of $MeV\!-$scale antineutrinos.
The high flux and penetrative properties of these weakly interacting particles have made reactors attractive sources for studying the fundamental properties of neutrinos, such as  oscillations and the neutrino mass hierachy~\cite{PhysRevD.78.071302,JUNO}. 
The same properties open the prospect for nonintrusive detection, exclusion, and cooperative monitoring of operating nuclear reactors at  tens of kilometers standoff, which are all applications of potential interest for nonproliferation~\cite{RevModPhys.92.011003}.

Kiloton to multikiloton-scale water and scintillator detectors, instrumented with thousands of photomultiplier tubes (PMTs), have been proposed for use in standoff reactor monitoring applications~\cite{askins2015physics}, and are now in use or planned for a range of fundamental physics studies with reactor sources~\cite{JUNO, derbin2016main, thekamlandcollaboration2013reactor}. 
In addition, a hundred kiloton--scale gadolinium-doped water detector known as Hyper-Kamiokande~\cite{HyperK} is under construction in Japan.
Both for nonproliferation applications~\cite{,akindele2022arms} and for basic physics studies (including possible DUNE Phase II detectors~\cite{DUNEWbLS}), interest exists from the international community in similarly large detectors filled with water-scintillator mixtures~\cite{alonso2014advanced}. 
These detector concepts take advantage of the size scaling properties of water while exploiting the advantages arising from the enhanced light output provided by the scintillating additive. 
 
The rate of signal interactions scales with detector volume.
Backgrounds also scale with volume and detector surface area, and must be mitigated. 
Purification of detector media, local shielding of gamma-rays and neutrons, and the use of overburden are all viable background suppression strategies, but these approaches increase the cost and complexity of deployment and operation. 
Additional  improvements in background rejection, reduced overburden and shielding, and other advantages might in principle be achievable without physical changes to the detector via improvements in the specificity of algorithms used to select physics signal events of interest within acquired data. 

A common approach to data selection is the development of physics-based selection criteria, using reconstructed energy, position, and other event features based on the known or expected properties of interacting particles. 
Distributions of the event features are examined to identify subsets preferentially associated with signal-like events. 
Selections of desired subsets of events from the feature phase space are then made to reduce the fraction of undesirable background events in the data set.  
This approach is commonly known as the `cut-and-count' method.  
Increasingly, data analysts have sought to augment or replace this approach by employing a range of machine learning (ML)--based methods to isolate event samples of interest. 

Compared to cut-and-count methods, ML-based approaches have the potential to improve sample purity, reduce computational effort, improve throughput in high data-rate experiments, and reduce biases arising in the formation and selection of reconstructed variables. 
Potential disadvantages include difficulty in understanding the underlying logic of ML-based selections and biases created by ambiguously defined or underpopulated training datasets.  
In this article, we compare two ML-based methods for extraction of simulated reactor antineutrino interactions from  backgrounds in gadolinium-doped water and water-based scintillator detectors. 

For specificity, we assumed a detector deployment site at the Boulby mine in Northern England~\cite{Boulby}. 
The mine is 26~km from two  nuclear reactors at the Hartlepool site,  each with a thermal power of $\!\!{~\sim}1500$ GigaWatts and a known and standard fuel loading. 
This site was proposed by the WATCHMAN collaboration as a possible location for a remote reactor monitoring demonstration\footnote{A 2021 announcement of an early shutdown of the Hartlepool reactors made the site less attractive for this demonstration.}. 
The reactor power and fuel loading allow us to predict the emitted antineutrino flux impinging on the detector. 
Muogenic neutrons, which form the principal backgrounds of interest for this study, are assumed to have a rate corresponding to the 1.1~km (2800 meters water equivalent) depth of the assumed underground deployment site at Boulby.  
Further details of the configuration and the detector are available in a 2019 Conceptual Design Report~\cite{NEO-CDR}. 
Simulated data were made available by members of the WATCHMAN collaboration, and generated using the RATPAC software package~\cite{RATPAC}.

The metric used to compare the effectiveness of the analysis techniques is relevant for nonproliferation applications. 
It is defined as the dwell time required to reject a null hypothesis with 3~$\sigma$ (about 99.7\%) confidence. 
We note that this standard is more stringent than the typical target statistical significance of $90\%$  used by the International Atomic Energy Agency in its fuel cycle monitoring activities~\cite{IAEAgloss}.

We consider two null hypotheses. 
Either no reactors are operational, or a single reactor is operating at its declared power throughout the observation period. 
The hypothesis is rejected if the measured rate of antineutrinos is at least three standard deviations above the expected background-only rate. 
Signal and background statistical fluctuations and systematic uncertainties were accounted for in the analysis.

We examine two ML-based approaches for antineutrino signal selection. 
A variant of the first approach has been studied in a recent article \cite{PRA_kneale_2023}, using the same simulated data and experimental configuration examined here. 
In that study, ML algorithms used reconstructed variables such as energy and position as inputs for training. 
The technique was shown to offer some improvement in signal to background ratio compared to standard cut-and-count methods.  
We additionally present an analysis that uses raw digitized time and amplitude signals output by the PMT array. 

Both ML methods improve detector sensitivity compared to cut-and-count methods. 
The technique using reconstructed data resulted in the greatest improvement in sensitivity. 
However, the performance of the raw-data approach can be expected to improve through the inclusion of additional underlying  physics information, notably the highly specific IBD positron-neutron time correlations. 
The raw-data method is also expected to reduce or eliminate biases that can appear in analyses which use  reconstructed variables as inputs, making it an attractive option for further application. 
Moreover, as other researchers have suggested~\cite{Cowen}, analysis of raw detector outputs may permit data selection algorithms to be imposed  early in the data-acquisition chain, prior to off-line reconstruction, possibly reducing data storage needs and/or increasing throughput of desirable events. 
We expect the raw-data approach to generalize best to other detectors using similar media and with similar design features, such as high channel counts and data rates.  

\section{Detector design}
The detector studied in this work is of a a kiloton-scale, liquid scintillator--based design for the detection of reactor antineutrinos via inverse beta decay. 
The design is based on the WATCHMAN detector concept~\cite{askins2015physics}, and our conclusions are relevant to similar kiloton-scale, multichannel PMT-readout detectors. 
This section considers the detector's design, operational principle, and the strategies employed to mitigate background signals, particularly from fast neutrons.

Two detector media are studied in this work: gadolinium-doped water (Gd-H$_2$O) and gadolinium-doped water-based liquid scintillator (Gd-WbLS). 
Gadolinium doping increases the neutron capture cross-section by several factors~\cite{PhysRevApplied.18.034059}. 
The percentage of Gd present in the two configurations  is 0.1\% by weight. 
WbLS is evaluated as a potential kton-scale detector medium owing to its higher scintillation yield, leading to better tagging efficiency for IBD signal events~\cite{Land2020}. 

\begin{figure}
    \centering
    \includegraphics[width=0.85\linewidth]{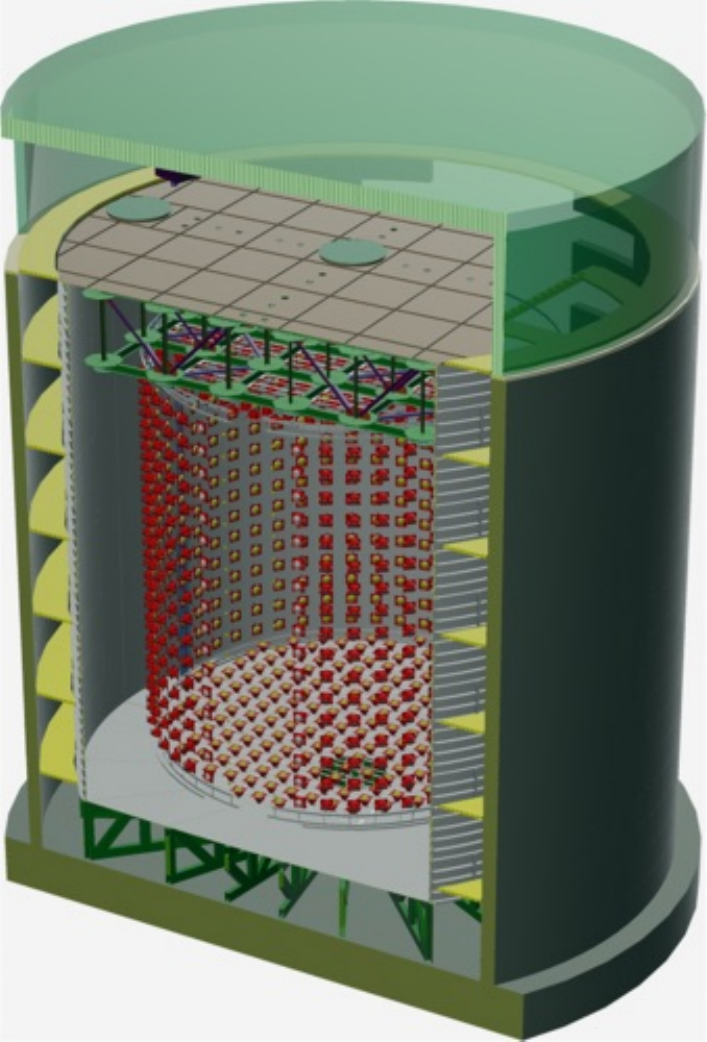}
    \caption{Visualization of the detector configuration created by Jan Boissevain (JG Boissevain Design). The inner cylinder is surrounded by PMTs, then an outer shield region of the same material.}
    \label{fig:detector}
\end{figure}

A 16~m diameter and height cylindrical tank houses the detector medium, including an outer veto/shield region of 2.5~m thickness and 226 outer-facing photomultiplier tubes (PMTs) to reject cosmogenic events.
The detector design is depicted in Fig.~\ref{fig:detector}.
Facing the inner active region of 11.4~m height and diameter are 2,330 PMTs (approximately 20\% photocoverage) and 1,232 PMTs (approximately 10\% photocoverage,) respectively, for the Gd-H$_2$O and Gd-WbLS configurations studied. 
Fewer PMTs are required for the Gd-WbLS fill due to its higher scintillation yield. 

In IBD, an antineutrino interacts with a quasi-free proton, producing a neutron and positron: 
\begin{equation}
    \bar{\nu}_e + p \rightarrow n + e^+  \ .
\end{equation}
At the MeV energy scales of reactor antineutrinos, the positron produces Cherenkov light in Gd-H2O, or a combination of scintillation and Cherenkov light in Gd-WbLS as it travels through the medium. 
The neutron, by contrast, is thermalized and ultimately captured within a few centimeters of its production.  
Most captures occur on Gd nuclei, due to the remarkably high capture cross-section of several Gd isotopes. 
The excited Gd nucleus decays rapidly, emitting gamma-rays isotropically with a total energy of $\sim$8~MeV~\cite{Hagiwara2018}.
The gamma-rays induce a combination of Cherenkov and scintillation light, forming a second, time-correlated signal associated with the IBD interaction.
The average delay between the positron and neutron signals is roughly 30~$\mu$s, and there is a strong positional correlation between the two signals in the detector. 

Some of the photons strike the PMT surfaces and create electronic pulses that can be clustered together to reconstruct an event.
The two correlated events in the detector from the positron and captured neutron allow for a signature pair of correlated events which are highly specific to IBD interactions. 
Based on their energy, time correlations and proximity, these event pairs can be tagged and backgrounds rejected for analysis. 

There are a host of backgrounds which can produce an event pair in the detector, ranging from accidental coincidences of uncorrelated events, cosmogenically activated long-lived isotopes that can create correlated background events, and muon-induced fast neutrons in the surrounding rock~\cite{fluka}. 
Muogenic fast neutrons are particularly challenging to reject using traditional background mitigation strategies. 
First, as penetrating neutral particles, they do not reliably trigger the veto system. 
Second, those escaping the veto and arriving in the main detector volume may generate two spallation neutrons, whose capture energies  are observable in the detector volume. 
This pair of neutron captures can mimic an IBD interaction due to the similar deposited energy and similar time and space correlations. 

Neutral products of muogenic neutrons do not produce Cherenkov-inducing positrons as IBD events do, offering an interesting opportunity to tag the IBD events and reject neutron-induced backgrounds according to their individual event characteristics, in addition to the correlation between events in a pair. 
Thus far, the primary method studied for reducing these IBD-like backgrounds has been by sequential selections on reconstructed high-level variable analogues of physical variables.  
For instance, a typical analysis might select based on the reconstructed event vertex, followed by the time delay until the second event, followed by the total signal amplitude seen in all PMTs for the first event (correlated to energy of the event), and so on. 
These methods are not inherently optimized to the multivariate task at hand, nor to identifying the combination of variables or data representation which provides the greatest separation between background and signal. 
To address this issue, we  reframe IBD signal selection as a machine learning problem, considering both high-level reconstructed event pair variables as one input representation to an ML algorithm, and for comparison, a raw-data approach to IBD signal classification treating only the first event in each event pair of signal and background data. 
These methods are presented in the following section, followed by a comparison of their sensitivity to IBD detection in Sec.~\ref{sec:analysis}.

\section{Event classification with machine learning}\label{sec:classification}
\subsection{Description of simulated data}

Particles were simulated using the AIT-NEO version of RATPAC (Reactor Analysis Tool Plus Additional Code)~\cite{RATPAC}. 
This software package tracks photon emissions arising from particle interactions, and  the collection of the generated photons by the photomultiplier tubes. 
The spectrum of muon-induced fast neutrons from the surrounding rock material was estimated using the Mei-Hime model~\cite{MeiAndHime}. 
The spectrum and production rates of IBD reactor events was obtained from the website geoneutrinos.org~\cite{geoneutrinos_paper}.

For each PMT, photon arrival times and the induced anode charge  are then used to reconstruct the original particle's vertex and direction, using a package called BONSAI~\cite{BONSAI} that was combined with AIT-NEO relevant functions. 
This combination of BONSAI and AIT-NEO relevant functions was wrapped into a private package called Functions for Event Reconstruction in Detector (FRED).
This simulation and processing software stack includes realistic event trigger conditions, event position reconstruction to which fiducial volume selection criteria are applied, evaluation of inner event distances, and calculation of the light isotropy.

After simulation and reconstruction, event pairs were identified by grouping events that were recorded within 600~$\mu$s of each other and isolated from all other events in the data by $1$~ms. 
This focused the analysis to consider only pairs of events which mimic IBD positron-neutron capture event pairs in their temporal signature, which are hardest to classify.

After these preprocessing and selection steps, there were approximately 10,322 event pairs comprising the Gd-H$_2$O simulated data and 13,608 event pairs comprising the Gd-WbLS simulated data, with each dataset containing an equal number of IBD signal event pairs as fast-neutron event pairs. 
Each dataset was partitioned into 75\% training and 25\% evaluation subsets.

\subsection{Tree-based classifier method}\label{sec:bf_class}

The first algorithm employed was a boosted decision tree model, enhanced in a novel way by using random forests as the individual base classifiers to reduce variance, and using the \verb|AdaBoost| boosting algorithm to reduce bias~\cite{Tu2017AdaBoost}. 
The model was developed and trained using the \verb|SciPy| software~\cite{scipy, scipy_paper}, and we will refer to it in discussion as the ``boosted forest" model.

\begin{figure*}
    \centering
    \includegraphics[width=\textwidth]{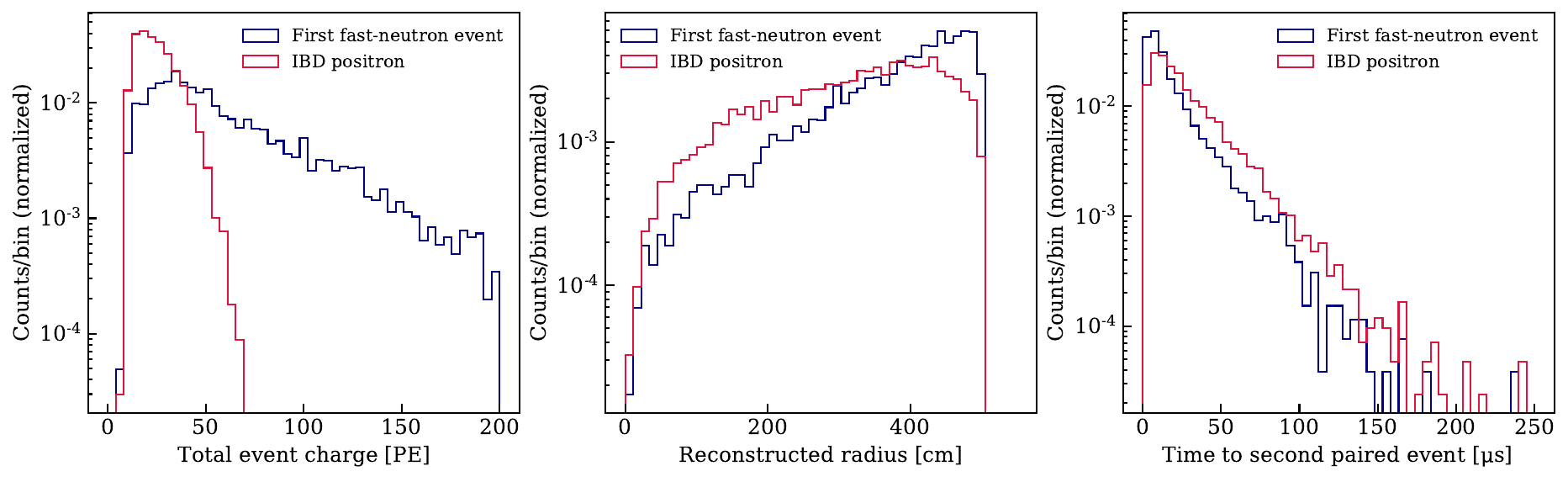}
    \caption{Reconstructed variables for IBD signal positron event and the first event from a fast-neutron induced event pair. The variables correspond to reconstructed energy (left), radial distance from the center axis of the detector (middle), and time between the first and second events in an event paid (right). The distributions illustrate the difficulty with background rejection based on reconstructed physical quantities.}
    \label{fig:fred_variables}
\end{figure*}

The reconstructed features of each event pair for signal and background data considered in the model were: 
information about the prompt and total PMT charge recorded during each event, their reconstructed positions, the time and distance between first and second event, the goodness-of-fit of the position and direction reconstruction of each event, and lastly variables to quantify the isotropy of the first event (namely the azimuthal KS-test and the first six average Legendre polynomial coefficients of cosine angle between PMT hits, formulated in~\cite{PhysRevLett.92.181301}). 
Key features are shown in Fig.~\ref{fig:fred_variables} to illustrate the class distributions upon these variables and the need for a multivariate, optimized approach to signal selection and background minimization. 

The model parameters were optimized using iterative scanning of overall accuracy.
The optimal parameters for Gd-H$_2$O (Gd-WbLS, in corresponding parentheses) were: 50 (15) random forest base classifiers, 50 (20) trees per random forest, each with a max depth of 8 (6) and a minimum number per final decision node of 10 (5). 
The last two parameters reduced overfitting given the limited training statistics. 

The performance on test data of the boosted forest models for binary classification of fast neutron-induced background vs. IBD signals is given in Table~\ref{table:bf_confusion}.
The results indicate that the boosted forest algorithm is a highly effective classifier for considering fast neutron background reduction, but it requires event-pair information at a reconstructed level, meaning that it is subject to a number of reconstruction steps and inter-event dependencies. 
Recently, efforts have been made to engineer novel machine learning applications to lower-level data to reduce reconstruction bias and improve signal selection efficiency~\cite{Rossi:2021tjf, XENONCollaboration:2023dar}. 
As such, we explored reconstruction-free, single event classification algorithms to illustrate the potential of such methods for kiloton-scale neutrino detectors.  

\begin{table}
\begin{center}
\begin{tabular}{ |c|c|c|} 
\multicolumn{3}{c}{Gd-WbLS} \\
\hline
& True IBD & True $n$ \\
\hline
Predicted IBD & 95.1~$\pm~1.5~\%$& 7.7~$\pm~2.1~\%$ \\ 
Predicted $n$ & 4.9~$\pm~1.5~\%$& 92.3~$\pm~2.1~\%$ \\ 
\hline 
\multicolumn{3}{c}{} \\
\multicolumn{3}{c}{Gd-H\textsubscript{2}O} \\
\hline
& True IBD & True $n$ \\
\hline
Predicted IBD & 93.4~$\pm~1.4~\%$& 11.5~$\pm~1.9~\%$ \\ 
Predicted $n$ & 6.6~$\pm~1.4~\%$& 88.5~$\pm~1.9~\%$ \\ 
\hline
\end{tabular}
\caption{Binary classification results for the boosted forest algorithm, considering both the WbLS and H$_2$O detector medium options. 
}
\label{table:bf_confusion}
\end{center}
\end{table}

\subsection{Graph convolutional network method}

A graph convolutional network (GCN) was developed for single-event classification using raw PMT data (meaning prior to application of reconstruction algorithms)  as input. 
The architecture of a GCN is designed to process graph-structured data with user-defined connectivity between data elements. 
GCNs enable the effective integration of topological information inherent in the detector's design without requiring, as traditional convolutional neural networks do, a transformation of the detector's spatial properties that would artificially introduce discontinuities into a fully connected spatial detector. 
A description of GCNs and their use cases can be found in~\cite{Yu2019Graph-Revised}. 

In this study, the graph nodes contain information from each individual PMT, and the graph edges between nodes represent potential connections between PMTs. 
Two PMT nodes shared a connective edge, thereby denoting potential causal connection of their recorded light, if they were within a 5~m Euclidian distance of each other. 
The edge requirement between two PMTs was optimized empirically, informed by the maximum opening angle size of the Cherenkov light cone, depending on detector medium~\cite{Caravaca2020, Kaptanoglu2021}. 

There were five input dimensions of data: PMT hit time relative to the center reconstructed time vertex, PMT position in $x$, $y$, and $z$, and the PMT charge observed during the first event in pairs of events. 
The output of the model was a single classifier node trained upon whether the first event was from an IBD reaction (and therefore a positron,) or from a fast neutron--induced event (likely a neutron capture). 

\begin{figure*}[t]
    \centering
    \includegraphics[width=\textwidth]{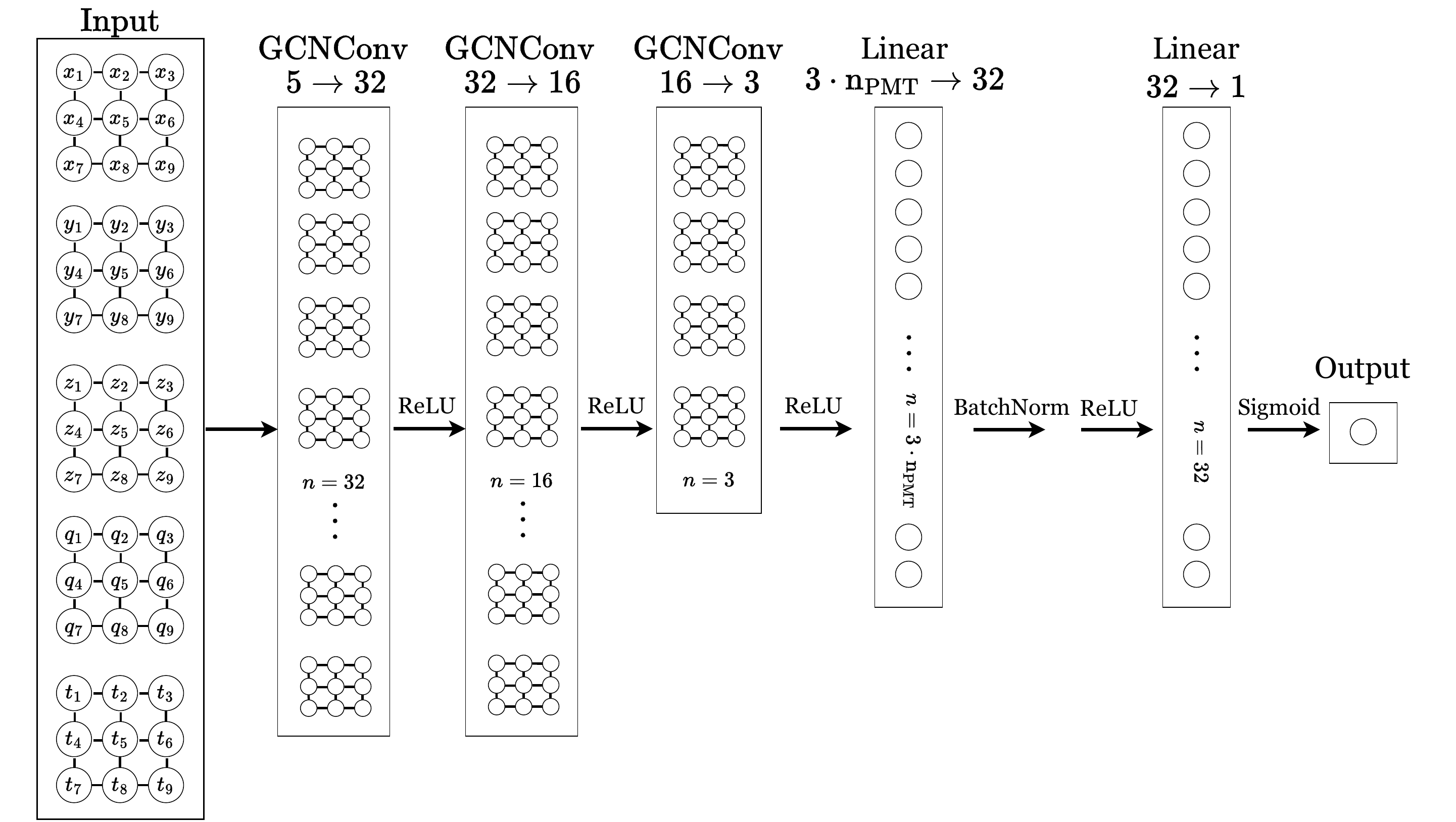}
    \caption{GCN architecture used for classifying first events in a pair as IBD-originating or fast neutron--originating. Three GCN convolutional filters comprise the heart of the model, with a final linear model applied to transform the graphical output to a standard single node for classification. The number of inputs in the graph are simplified for visualization, but correspond in reality to the number of PMTs in each detector configuration. The edge connectors between nodes in the graph represent that the two PMTs are within 5~m from each other. The input variables are the $x$, $y$, and $z$ positions of each PMT, the charge observed by each PMT, and the time of observed charge relative to the center reconstructed time of the event.}
    \label{fig:gcn_architecture}
\end{figure*}

A summary of the GCN architecture is shown in Fig.~\ref{fig:gcn_architecture}.
The architecture of the GCN was the same for the Gd-H$_2$O and Gd-WbLS dataset and PMT configurations, excepting only the number of nodes, which corresponded to the number of PMTs in each detector configuration. 
The GCN contained three graph convolutional layers: the first layer maps the 5-dimensional input feature space to a 32-dimensional hidden space, followed by a reduction to 16 dimensions in the second layer, and finally, the features are compressed to a 3-dimensional space in the third layer.
Each convolutional layer is followed by a ReLU activation function to introduce nonlinearity into the model, facilitating the learning of complex patterns within the data. 
The output from the last graph convolutional layer is reshaped and passed through a linear sequence of operations, including a linear transformation to 32 dimensions, batch normalization for stabilizing learning, another ReLU activation, and a final linear transformation to produce a one-dimensional output contains a sigmoid activation function. 
This output signifies the model's prediction regarding the classification of the input event.

Training was conducted on an NVIDIA GPU utilizing CUDA technology, leveraging the computational power necessary for processing the graph-structured data efficiently.
We utilized \verb|PyTorch|~\cite{pytorch} to implement our model architecture and training procedure.
Regarding additional model training parameters, the model was trained for 30 epochs, employing a learning rate of 3$\times$10$^{-3}$ and the \verb|Adam| optimizer option. 
The loss function employed was binary cross-entropy. 

The model was evaluated on the data test set identically in procedure to the boosted forest classifier. 
For binary classification performance, a decision boundary on the output node of 0.5 was used to perform basic classification and evaluation of the confusion matrix in Table~\ref{table:gcn_confusion}. 
The GCNs excelled at classifying IBD positrons accurately for each configuration, but relative to the boosted forest method, had greater confusion in correctly classifying the fast neutron--induced first event. 

We hypothesize that this is due to the fact that, in some cases, fast neutrons can scatter and create a  delta-ray, a high energy electron that can produce a few~$MeV$ Cherenkov signal, an energy similar to that of an IBD-generated positron. 
To confirm this, we studied the distribution of the beta coefficients of the Legendre polynomials described in Sec.~\ref{sec:bf_class} most reflective of anisotropy in the PMT signals for the first events in both the fast neutron and IBD cases.
We found that the errors in the trained GCN for fast neutron events followed a distribution of these beta variables that deviated from the rest of the fast neutron training sample, yet agreed with the distribution of that of the IBD positrons, supporting our understanding that these events originated from simulated delta-ray emissions arising from fast neutron interactions. 
This result indicates that the GCN learned the unique Cherenkov or charged-signal-like signature in the PMTs.
In the future, event-by-event analysis of this sort could be a useful output variable to include in individual event topology studies. 

The GCN approach offers distinct advantages and future applications as an analysis tool for kiloton-scale neutrino detectors which should be considered.
Specifically, the model operates directly on raw data, bypassing dependence on feature extraction and preprocessing for event selection. 
This approach not only simplifies the data pipeline but also provides an additional metric of interest likely linked to Cherenkov anisotropy. 
This variable could provide valuable insights into the physical processes underlying the detected events, emphasizing the potential of the GCN to contribute to a deeper understanding of the phenomena being observed in a wide host of physical processes beyond fast neutron--induced events. 

Having demonstrated the fundamental knowledge learned by this classification technique, a full-scale production of high statistics simulations of signal and background data is merited. 
Despite the $\sim$10\% reduction in true positives in the raw performance metrics by the GCN, the ability to analyze events based on raw detector signals presents a compelling case for using GCNs to study novel characteristics of neutrino and cosmogenic interactions in these types of detectors. 

\begin{table}
\begin{center}
\begin{tabular}{ |c|c|c|} 
\multicolumn{3}{c}{Gd-WbLS} \\
\hline
& True IBD & True $n$ \\
\hline
Predicted IBD & 82.9~$\pm~2.6~\%$& 22.9~$\pm~2.9~\%$ \\ 
Predicted $n$ & 17.1~$\pm~2.6~\%$& 78.1~$\pm~2.9~\%$ \\ 
\hline 
\multicolumn{3}{c}{} \\
\multicolumn{3}{c}{Gd-H\textsubscript{2}O} \\
\hline
& True IBD & True $n$ \\
\hline
Predicted IBD & 80.9~$\pm~2.3~\%$& 40.4~$\pm~2.7~\%$ \\ 
Predicted $n$ & 19.1~$\pm~2.3~\%$& 59.6~$\pm~2.7~\%$ \\ 
\hline
\end{tabular}
\caption{Binary classification results for the GCN algorithm, considering both the WbLS and H$_2$O detector medium options. 
}
\label{table:gcn_confusion}
\end{center}
\end{table}

\section{Analysis of sensitivity to reactor signals}\label{sec:analysis}

The two trained models were then evaluated according to their impact on the sensitivity of the detector to reactor antineutrinos by calculating the dwell time to measurement described in Sec.~\ref{sec:introduction}.
Available to us in this case were the outputs of the two ML models, which we term the ``scores": the boosted forest classification prediction averaged over all ensemble members, and the output of the final node of the GCN. 
A decision boundary upon the scores of each model was used to select events which pass the ML criterion and are included in the final observed event rate. 

The score boundaries were converted into the relative fractions of signal and background events accepted at a given decision boundary on the score.
These fractions then were converted into absolute rates using the standard rates of signals and backgrounds from all background sources. 
The dwell time for reactor measurement to significance, $Z$, is evaluated using a Profile Likelihood method~\cite{LUNDBERG2010683}.
The likelihood when a source of signal is expected in the presence of a known background rate is,
\begin{widetext}
\begin{eqnarray}
    L (\mu, b, \epsilon|x,\epsilon_o,b_o)&=& P_{poisson}(x|\mu,\epsilon,b)\cdot P_{gauss}(b|b_o,\sigma_b)\cdot P_{gauss}(\epsilon|\epsilon_o,\sigma_\epsilon)   \ ,
\end{eqnarray}
\label{Likeli}
\end{widetext}
where $P_{gauss}$ and $P_{poisson}$ are the gaussian and poisson distributions, $\mu$ is the expected antineutrino-induced positron rate in the detector, $\epsilon$ is the signal detection efficiency, $b$ is the total background rate from all sources, and $\sigma$ is the uncertainty for each rate. 
Here $b_o$ is the expected background rate, which can be estimated using Monte Carlo, calibration data or other methods, and $\epsilon_o$ is the nominal detection efficiency. In our analysis later in the paper, note the signal rate $s$  is defined as $s\equiv\mu\epsilon$.  
The $-2\log(L)$ term defining $Z$ is similar to a $\chi^2$ distribution with one degree of freedom. 
While three variables are present: $\mu$, $\epsilon$, and $b$,
the latter two are nuisance parameters.

The advantage of the profile likelihood method is that it offers a way to treat nuisance parameters such as uncertainties on signal and backgrounds using a single significance metric. 
The likelihood ratio value $\lambda$ is given by,
\begin{equation}\label{eq:ratio}
    \lambda(\mu) = \frac{\sup L(\mu,\hat{b}(\mu),\hat{\epsilon}(\mu)|x,b_o,\epsilon_o)}{\sup L(\hat{\mu},\hat{b},\hat{\epsilon}|x,b_o,\epsilon_o)}\ ,
\end{equation}
where $\hat{b}(\mu)$ and $\hat{\epsilon}(\mu)$ are the supremum values of equation \ref{Likeli}.  
We note that $\log(\sup L)$ leads to the same solutions as $\sup \log L$, leading to the sensitivity metric,
\begin{equation}
    Z = \sqrt{-2\log\lambda(\mu)}\ .
\end{equation}

Evaluating the denominator of Eq.~\ref{eq:ratio} leads to values of $\hat{\mu}=(x-b_o)/\epsilon_o$ and $\hat{b}=b_o$ and $\hat{\epsilon}=\epsilon_o$. 
For the numerator of Eq.~\ref{eq:ratio}, the values of $\hat{b}(\mu)$ and $\hat{\epsilon}(\mu)$ can be solved analytically, and solutions are available~\cite{LUNDBERG2010683}.

Until an experiment takes data, we cannot know what the number of observed events $x$ will be. 
Nonetheless, two scenarios can be useful to predict the sensitivity to certain questions relevant to the non-proliferation community. 
The are two cases of interest: a maximal signal limit for the case when no clear signal is expected, and the case where a signal is expected and observed at a certain confidence level, also known as the the experimental sensitivity case. 

The no clear signal case is simply when the number of observed events is equal to the average expected background ($x=b_o$) and the experimental sensitivity case is when the number of observed events is the average expected signal plus average background ($x=\epsilon_o\mu+b_o$), which is also known as the Asimov data set. 
The Profile Likelihood offers good coverage of both of these options.

Background rates were considered for the following components: radiogenic backgrounds including from long-lived isotopes and fast neutrons, and background IBD events from world reactors and geoneutrinos. 
Accidental coincidence rates are negligible in this study due to the data preselection algorithm's effectiveness at reducing accidental rates. 
Signal rates considered two reactor operation scenarios: core 1 of the reactor on and known, while core 2 turns on (and is treated as the "signal" core), which we call the 1-core scenario, and both core 1 and 2 turned on as the "signal" of antineutrinos (which we call the 2-core scenario). 
The values of rates studied here are given in Table~\ref{table:rates} for the Gd-H$_2$O and Gd-WbLS detector configurations\footnote{ This hypothetical experimental scenario is a demonstration using a pair of high-power reactors. The dwell-times calculated are thus not reflective of a real-world reactor exclusion or discovery scenario, which would more likely involve exclusion of relatively low-power reactors in areas without nearly operating reactors.}.

\begin{table*}[htb]
  \centering
  \renewcommand{\arraystretch}{1.7}
  \setlength{\tabcolsep}{10pt} 

  \begin{tabularx}{0.778\textwidth}{|>{\hsize=2.5\hsize}X |*{5}{c|} @{}}
    \multicolumn{1}{c}{} & \multicolumn{4}{c}{\textbf{Component rates}} \\
    \hline
    \multicolumn{1}{| c|}{\textbf{Configuration}} & Core 1 / Core 2 & Fast neutrons & IBD BG & Radiogenics  \\
    \hline
     \multicolumn{1}{| c|}{Gd-H\textsubscript{2}O} & 0.397 $\pm$ 0.035 & 0.237 $\pm$ 0.115 & 0.013 $\pm$ 0.002 & 0.059 $\pm$ 0.034 \\
     \multicolumn{1}{| c|}{Gd-WbLS} & 0.405  $\pm$ 0.036  & 0.160 $\pm$ 0.078& 0.020 $\pm$ 0.004 & 0.033 $\pm$ 0.019\\
    \hline
  \end{tabularx}
  \caption{Rates, in events per day, of various physical sources after the standard preselection algorithm has been applied. These rates are used to calculate the ML-informed rates at a given output acceptance values for the ML algorithms studied. Two sources of systematic uncertainties have been published in the literature~\cite{Akindele_uncertainties,PRA_kneale_2023}. For these studies the more conservative relative systematic uncertainties of~\cite{Akindele_uncertainties} were used, and applied to both the Gd-H$_2$O and Gd-WbLS cases. Additionally, early studies showed that accidental backgrounds were almost all removed ($>$99\%) through the ML techniques described in the paper~\cite{PRA_kneale_2023}and are not considered for this study.}
  \label{table:rates}
\end{table*}

The dominant contribution to the background rate comes from fast neutrons.
We conservatively assume that the subdominant contributions to the background rates arising from radionuclides are not reduced by the ML algorithms, and that background IBDs are accepted at the same fraction as signal IBDs.  

The comparison baseline method of signal selection is by selecting on a few reconstructed variables of interest to reduce background rates. 
These correspond to the prior data selection before the ML algorithm was trained or evaluated, enforcing a general prior selection on event pairs within the energy and volume regions of interest. 
These are customizable preprocessing options in analysis that in the future can be optimized in tandem with the optimal decision boundary on the ML method. 

\begin{figure*}[t]
    \centering
    \includegraphics[width=\textwidth]{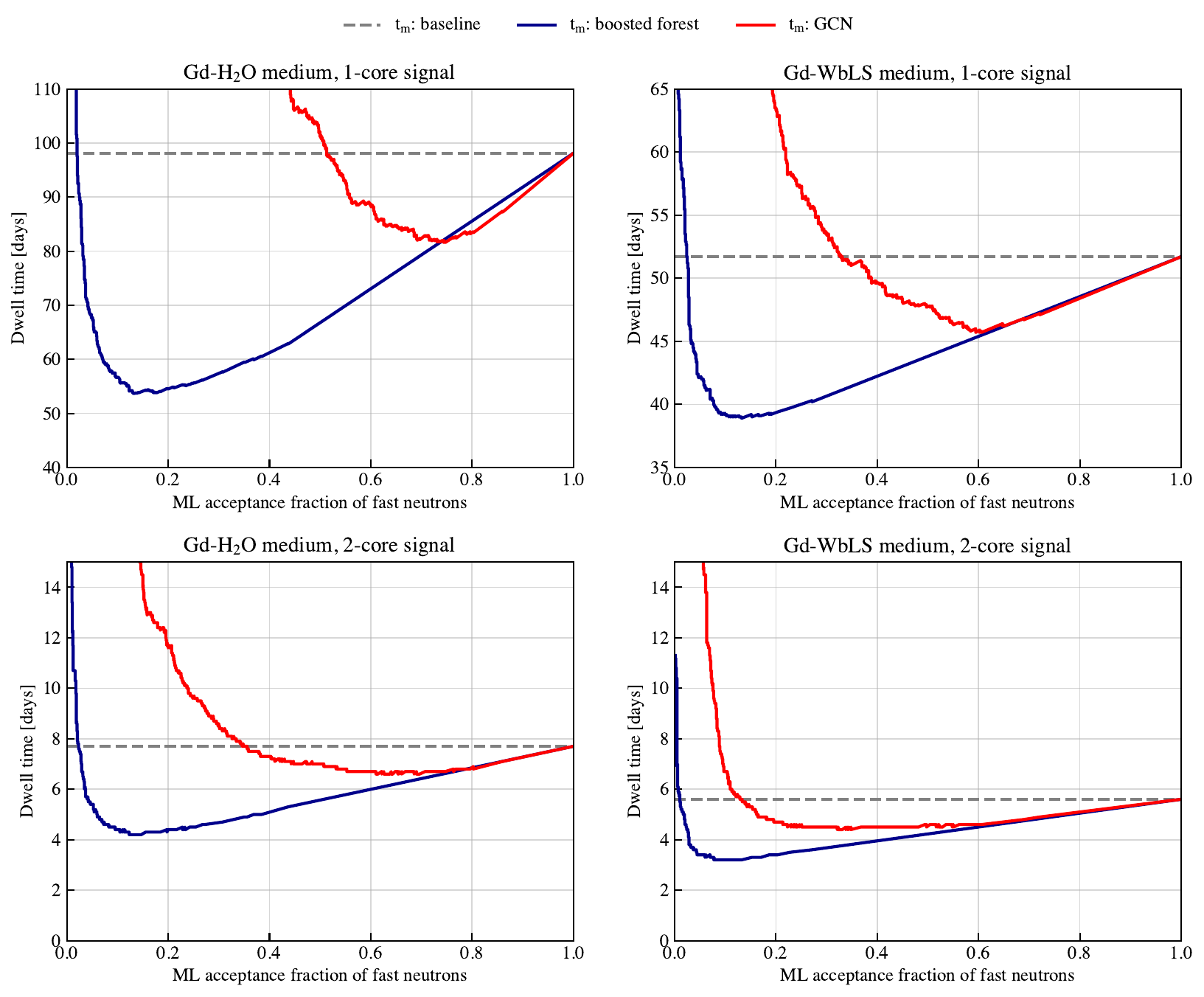}
    \caption{Dwell times as a function of the ML-driven acceptance fraction of fast-neutron backgrounds, for various configurations. 
    Top: reactor configured in 1-core operating mode (only core 1 in Table~\ref{table:rates} considered as signal rate); Bottom: reactor configured in dual-core operating mode (both signal rates in Table~\ref{table:rates} considered); Left: Gd-H$_2$O detector configuration; Right: Gd-WbLS detector configuration. 
    Dashed lines show as reference the value of the measurement dwell times ($t_{\text{m}}$) as obtained in a standard analysis approach.
    In each configuration, both the boosted forest and GCN methods of ML-driven signal selection result in improved measurement dwell times. 
    }
    \label{fig:dwell_time}
\end{figure*}

The signal and background acceptances were translated into dwell times as a function of the fraction of total background accepted, which are depicted in Fig.~\ref{fig:dwell_time} for the various detector configurations and reactor operation scenarios. 
These functions were then used to determine the optimized dwell times of the various methods, which are reported in Table~\ref{table:dwell_opt}. 
The results show that for both ML methods studied, there is an optimal acceptance boundary on the model output, corresponding to an optimal fractional background rate, which is dependent on the ML method in question and the detector and reactor configurations. 

\begin{table*}[ht]
  \centering
  \renewcommand{\arraystretch}{1.3} 
  \setlength{\tabcolsep}{10pt} 
  \begin{tabularx}{.62\textwidth}{|>{\hsize=7\hsize}X | *{2}{c|} *{2}{c|} }
    \multicolumn{1}{c}{} & \multicolumn{4}{c}{\textbf{Measurement dwell time, $t_m$}}  \\
    \hline
    \multicolumn{1}{|c|}{} & \multicolumn{2}{c|}{\textbf{1-core signal}} & \multicolumn{2}{c|}{\textbf{2-core signal}}  \\
    \multicolumn{1}{|c |}{\textbf{Method}} & \mbox{Gd-H$_2$O} & \mbox{Gd-WbLS} & \mbox{Gd-H$_2$O}  & \mbox{Gd-WbLS}  \\
    \hline
    \multicolumn{1}{|c |}{Baseline} & 98.1 & 51.7 & 7.7 & 5.6 \\
    \multicolumn{1}{|c |}{Boosted Forest} & 53.7 & 38.9 & 4.2  & 3.2 \\
    \multicolumn{1}{|c |}{GCN} & 81.6 &  45.7 &  6.6 & 4.4 \\
    \hline
  \end{tabularx}
  \caption{Optimized dwell times in days for the studied analysis methods. In all configurations, the shortest dwell time is obtained by use of the boosted forest method of event selection.}
  \label{table:dwell_opt}
\end{table*}

Across all configurations studied, the boosted forest algorithm had the best performance.
This likely reflects the use of the interevent time information in this decision model. In addition, the nature of the data representation given the limited statistics in this case tended to favor the reconstructed variable-based decision tree model. 
The GCN method still showed improvement over the baseline selection method, even though it was only being trained and evaluated upon the first event in each event pair. Like the boosted forest algorithm, the GCN approach would likely benefit from the inclusion of training on interevent time patterns, which typically have high rejection power in cut-and-count analyses. 
Additionally, the GCN output was found to be informative beyond classification into event topology based solely on the PMT signals of the first event, emphasizing its potential for characterizing the underlying physical process of background events. 

\section{Conclusion}

This study presents the first comparative analysis between a boosted random forest model applied to reconstructed event pair data and a graph convolutional network (GCN) model processing raw PMT data in the context of water-based Cherenkov detectors for nuclear reactor monitoring. 
Our findings underscore the significant potential of ML methodologies to enhance traditional cut-and-count approaches in detecting reactor antineutrinos, enabling new detector configurations and experimental operations by improving the sensitivity of these rare-event detectors. 

Both the boosted forest and GCN models demonstrated superior detection sensitivity compared to standard analytical methods across all detector configurations, reactor operating scenarios, and metrics of sensitivity considered. 
This improvement is particularly noteworthy given the models' abilities to handle complex background noise and signal overlap, a common challenge in neutrino detection.
The adaptability of these ML methods to different detector and reactor operational conditions highlights both their robustness and their wide applicability for kiloton-scale water-based neutrino detectors. 
We expect that similar strategies can be employed profitably in other large-scale neutrino experiments, such as the DUNE Module of Opportunity, JUNO, and SuperKamiokande. 

A few key areas merit future investigation. 
A comprehensive simulation campaign tailored to these networks and considering all background sources would be instrumental in improving the precision and reliability of neutrino detection.
The development of simulation and processing techniques, for instance through generative neural networks~\cite{Musella:2018rdi}, could lead to more sophisticated and robust models that are capable of extracting subtle patterns and correlations within the data.
For instance, developing a geometric neural network~\cite{SHANG2021104364} upon the results of our GCN for the optimization and reconstruction of critical parameters, such as the anisotropy factor of signals observed in an event. 
Incorporating Bayesian networks~\cite{Pogwizd:2007sp, XENONCollaboration:2023dar} to develop probabilistic signal/background metrics represents another exciting frontier. 
This approach would enable a more nuanced analysis of the data, leveraging probabilistic information to refine the distinction between signal and background events.
Finally, the GCN approach, being independent of reconstructed variables, is well-matched for possible implementation in the early stages of an online triggering system, such as within a Field-Programmable Gate Array accepting raw input signals from the PMTs. 
Such an approach to data acquisition could bring advantages in terms of throughput of physics events of interest, and possibly significant reduction in the volume of stored event data. 

In summary, this work establishes that machine learning models can enhance antineutrino detection in kiloton-scale water-based Cherenkov detectors. 
The demonstrated improvement in background rejection can be used in practical ways to simplify designs of detectors for remote reactor monitoring, for example by improving performance at shallower overburden due to the ability to reject higher fractions of muogenic backgrounds. 
We anticipate that similar performance advantages will inhere for fundamental physics experiments involving reactor antineutrinos, and we expect the performance gains observed here will be broadly applicable for other physics goals in similar water/scintillator detectors.  
By harnessing the power of advanced ML techniques, future efforts can continue to push detection boundaries in neutrino physics, contributing to our understanding of fundamental physics and bolstering global nuclear nonproliferation efforts.

\section*{Acknowledgments}
Part of this work was performed under the auspices of the U.S. Department of Energy by Lawrence Livermore National Laboratory under Contract DE-AC52-07NA27344,
LLNL-JRNL-865846. S. Farrell was supported through DOE grant NE-NA0003960.
Computations were performed on the Livermore-Computing Lassen cluster. 
We thank the WATCHMAN collaboration for useful comments and critiques at various stages of this study.
We  express our gratitude to Ron Wurtz and Brenton Blair, formerly of LLNL, for their useful insights in the initial definition and implementation of this work, and to Aaron Higuera of Rice University for his helpful advice throughout the study.
We also  thank  Jan Boissevain (J.G. Boissevain Design) for providing  Fig.~\ref{fig:detector}.

\bibliography{main.bib}
\end{document}

%% file: affiliations.tex
\newcommand{\rice}{\affiliation{Department of Physics and Astronomy, Rice University, Houston, TX 77005, USA}}
\newcommand{\llnl}{\affiliation{Lawrence Livermore National Laboratory, Livermore, CA 94550, USA}}
\newcommand{\ab}{\affiliation{Vienna Center for Disarmament and Nonproliferation, Vienna, Austria}}

%% file: authors.tex
\author{S.~Farrell}\email[]{drfarrellconsulting@gmail.com}\rice
\author{M.~Bergevin}\llnl
\author{A.~Bernstein}\ab